\documentclass[preprint,11pt,numbers,sort&compress]{elsarticle}
\pdfoutput=1
\PassOptionsToPackage{table}{xcolor}
\usepackage{geometry}
\usepackage[T1]{fontenc}
\usepackage[utf8]{inputenc}
\usepackage{lmodern}
\usepackage{mathpazo}
\usepackage[varqu,varl,var0,scaled=0.97]{inconsolata}
\usepackage[colorlinks = true]{hyperref}
\usepackage{amsmath,amsfonts,amssymb}
\usepackage{url}
\makeatletter
\g@addto@macro{\UrlBreaks}{\UrlOrds}
\makeatother
\usepackage{mmacells}
\usepackage{mathtools}
\usepackage{xcolor}
\usepackage{slashed}
\usepackage{xspace}
\usepackage[absolute]{textpos} 

\journal{Computer Physics Communications}

\newcommand{\feyncalc}{\textsc{FeynCalc}\xspace}
\newcommand{\formcalc}{\textsc{FormCalc}\xspace}
\newcommand{\feynarts}{\textsc{FeynArts}\xspace}
\newcommand{\pax}{\textsc{Package-X}\xspace}
\newcommand{\fire}{\textsc{FIRE}\xspace}
\newcommand{\feynhelpers}{\textsc{FeynHelpers}\xspace}
\newcommand{\mma}{\textsc{Mathematica}\xspace}
\newcommand{\form}{\textsc{FORM}\xspace}

\DeclareMathOperator{\Tr}{Tr}
\newcommand*{\cf}{cf.\ }
\newcommand*{\eg}{e.\,g.\ }
\newcommand*{\ie}{i.\,e.\ }
\newcommand*{\Eq}{Eq.\,}
\newcommand*{\Sec}{Section\xspace}

\mmaSet{morefv={gobble=2},morelst={breaklines=true}}

\begin{document}
\begin{frontmatter}

\title{FeynCalc 9.3: New features and improvements}

\author[a,b,c]{Vladyslav Shtabovenko\texorpdfstring{\corref{author1}}{}}
\author[d]{Rolf Mertig\texorpdfstring{\corref{author2}}{}}
\author[e]{Frederik Orellana\texorpdfstring{\corref{author3}}{}}

\cortext[author1] {\textit{E-mail address:} v.shtabovenko@kit.edu}
\cortext[author2] {\textit{E-mail address:} rolfm@gluonvision.com}
\cortext[author3] {\textit{E-mail address:} fror@dtu.dk}

\address[a]{Institut f\"ur Theoretische Teilchenphysik (TTP), Karlsruher Institut
f\"ur Technologie (KIT), 76131 Karlsruhe, Germany}
\address[b]{Zhejiang Institute of Modern Physics, Department of Physics, Zhejiang University, Hangzhou, 310027, China}
\address[c]{Technische Universit\"at M\"unchen, Physik-Department T30f, James-Franck-Str. 1, 85747 Garching, Germany}
\address[d]{GluonVision GmbH, B\"otzowstr. 10, 10407 Berlin, Germany}
\address[e]{Technical University of Denmark, Anker Engelundsvej 1, Building 101A,
2800 Kgs. Lyngby, Denmark}

\begin{textblock*}{100ex}(0.8\textwidth,5ex)
{P3H-20-002, TTP19-020,  TUM-EFT 130/19}
\end{textblock*}

\begin{abstract}
We present \feyncalc 9.3, a new stable version of a powerful and versatile \mma package for symbolic quantum field theory (QFT) calculations. Some interesting new features such as highly improved interoperability with other packages, automatic extraction of the ultraviolet divergent parts of 1-loop integrals, support for amplitudes with Majorana fermions and $\gamma$-matrices with explicit Dirac indices are explained in detail. Furthermore, we discuss some common problems and misunderstandings that may arise in the daily usage of the package, providing explanations and workarounds.
\end{abstract}

\end{frontmatter}

\section{Introduction}

Feynman diagrammatic expansion is one of the most common techniques to approach perturbative calculations in modern theoretical high energy physics (HEP). For the given theory each Feynman diagram can be converted to an algebraic expression (\eg scattering amplitude) using a set of well-defined rules (Feynman rules). In principle, every single step of a typical HEP calculation from deriving Feynman rules for the given Lagrangian till obtaining explicit results for a particular observable can be carried out using computer codes. In practice, different tools are required to automatize different parts of the calculation and it is often far from trivial to interface and employ multiple codes in an efficient, correct and consistent way. Given the sheer multitude of HEP programs that are publicly available or can be obtained upon request by contacting the authors, it may not be easy to decide which tool fits best for one's project.

This is why we would like to attract the attention of the reader to \feyncalc~\cite{Mertig:1990an,Shtabovenko:2016sxi}, a tool that is particularly well suited for manipulating Feynman amplitudes in a specific way, where it is important to keep track of each calculational step. \feyncalc is an open-source \mma package that can be used standalone or conveniently integrated in a custom computational setup. It has been available to the HEP community for almost three decades and is still actively developed and widely used in various fields of particle physics, ranging from Standard Model (SM) and Beyond Standard Model (BSM) processes to Effective Field Theory (EFT) calculations.

Together with \pax~\cite{Patel:2015tea,Patel:2016fam} and \textsc{HepMath}~\cite{Wiebusch:2014qba}, \feyncalc belongs to the category of software tools for semi-automatic calculations. The main goal of such packages is to provide the user with an extensive toolbox for symbolic QFT calculations (including, but not limited to, the evaluation of Feynman diagrams), automatizing such common operations as index contractions, matrix chain simplifications, manipulations of loop integrals or summations over polarizations of external particles. These operations can be performed in any order according to the best judgment of the user. Indeed, the practical usage of \feyncalc is conceptually very similar to pen and paper calculations, with all the related merits and drawbacks. A skillful practitioner can employ the package in highly nontrivial ways, obtaining results that would be difficult to get using a more automatic but less flexible framework. Yet to avoid mistakes and wrong results it is absolutely necessary to possess a good understanding of technical subtleties and especially of the physics behind the particular calculation.

This design philosophy is very different from the behavior of more automatic publicly available frameworks such as \textsc{MadGraph5\-\_aMC@NLO}~\cite{Alwall:2014hca}, \textsc{GoSam}~\cite{Cullen:2011ac,Cullen:2014yla}, \textsc{Herwig++}~\cite{Bahr:2008pv}, \textsc{Sherpa}~\cite{Gleisberg:2008ta}, \textsc{Whizard}/\textsc{O'Mega}~\cite{Moretti:2001zz,Kilian:2007gr}, \feynarts/\formcalc~\cite{Hahn:1998yk,Hahn:2000kx}, \textsc{CalcHep}~\cite{Belyaev:2012qa}, \textsc{CompHep}~\cite{Boos:2004kh} and many others. Although here the user may somewhat influence the course of the calculation via different options and switches specified in the input card file, all the technicalities are taken care of by the automatic algorithms under the hood. While this approach is certainly well justified from the point of view of creating fast and easy-to-use codes, it also makes it rather difficult to compute anything that was not envisaged by the package developers or to compute things in a different way. 

Of course, the enormous flexibility of \feyncalc due to is its reliance on \mma for all algebraic manipulations comes at a price. The performance is admittedly much worse as compared to codes that use faster symbolic manipulation systems, notably \form~\cite{Vermaseren:2000nd,Kuipers:2012rf}. While this may not be too problematic for smaller calculations that finish within minutes on a modern computer,\footnote{An interesting approach to such projects is the new \textsc{FeynMaster}~\cite{Fontes:2019wqh} framework.} the matter becomes much more pressing when dealing with the increasing number of diagrams, loops and legs. Bearing in mind the intrinsic limitations of \mma when manipulating very large expressions with hundreds of thousands or even millions of terms, the only sensible way to address this issue is to adopt a hybrid approach,
by offloading computationally heavy operations to \form, while still employing \feyncalc for more appropriate tasks. This strategy was successfully applied \eg in the recent calculation of the Energy\--Energy\--Correlation function in gluon-initiated Higgs decays at NLO~\cite{Luo:2019nig}, where the most demanding parts of the computation (squaring of double-real matrix elements and reduction of 3-loop integrals) were done with the aid of \form, \textsc{color} ~\cite{vanRitbergen:1998pn} and \fire~\cite{Smirnov:2014hma}, while \feyncalc was used to work out linearly independent bases for loop integral topologies, recalculate the simpler amplitudes for real-virtual corrections and perform various cross-checks. In this sense we believe that it is perfectly possible to employ \feyncalc in complicated multi-loop setups as long as one does not try to calculate \textit{everything} relying solely on \feyncalc and \mma.

The aim of this paper is to report on the new \feyncalc 9.3, the next major release since 2016. The installation of the package is covered in \Sec\ref{sec:installation}, while \Sec\ref{sec:features} describes new features introduced in this version. In \Sec\ref{sec:usage} we provide best practices for using \feyncalc in loop calculations. The current status of the project and an outlook towards future developments are summarized in \Sec\ref{sec:summary}.

\section{Installation} \label{sec:installation}

\subsection{Automatic installation}

The most convenient way to install the current stable version of \feyncalc is to evaluate the following line
{\small
\begin{mmaCell}[index=1,moredefined={InstallFeynCalc}]{Input}
  Import@"https://raw.githubusercontent.com/FeynCalc/feyncalc/master/install.m"
  InstallFeynCalc[]
\end{mmaCell}
}
\noindent in a \mma (version 8 or later is required) notebook that runs on a new kernel. The automatic installer essentially downloads the current snapshot of the package from the \texttt{hotfix-stable} branch of the main repository on \textsc{GitHub}\footnote{\url{https://github.com/FeynCalc/feyncalc}} and sets up the diagram generator \feynarts to work with \feyncalc out-of-the-box. Since the stable version of \feyncalc receives bug fixes until a new official release, it is useful to reinstall the package from time to time to ensure that one is running the most recent version. The updates to the stable version are visible as commits to the \texttt{hotfix-stable} branch\footnote{\url{https://github.com/FeynCalc/feyncalc/commits/hotfix-stable}} of the repository.

The automatic installer may not work with \mma versions 8 and 9 due to their limitations when accessing URLs via HTTPS. In that case one can always download all the necessary files by hand and run the installer in the offline mode or proceed with a manual installation. The corresponding instructions including a collection of workarounds for older \mma versions can be found in the project wiki.\footnote{\url{https://github.com/FeynCalc/feyncalc/wiki/Installation}}

Users who employ the \textsc{Free Wolfram Engine for Developers} and use \textsc{Jupyter} as a front end should install \feyncalc via the text-based interface.\footnote{\url{https://reference.wolfram.com/language/tutorial/UsingATextBasedInterface.html}} Although \feyncalc can be loaded in a \textsc{Jupyter Notebook}, not all front end related features may work correctly due to the existing limitations of the \textsc{Jupyter} interface.
 
It should be noted that in \feyncalc 9.3 the internal mechanism to load \feynarts and \textsc{TARCER}~\cite{Mertig:1998vk} has been reorganized in
the form of add-ons so that instead of \texttt{\$Load\-FeynArts} and \texttt{\$Load\-TARCER} one should use
{\small
\begin{mmaCell}[moredefined={LoadAddOns, FeynCalc}]{Input}
  \$LoadAddOns=\{"FeynArts"\};
  <<FeynCalc`
\end{mmaCell}
}
\noindent and
{\small
\begin{mmaCell}[moredefined={LoadAddOns, FeynCalc}]{Input}
  \$LoadAddOns=\{"TARCER"\};
  <<FeynCalc`
\end{mmaCell}
}
\noindent respectively. The old syntax is still supported for compatibility reasons. Since \texttt{\$Load\-Add\-Ons} is a list of strings, one can equally specify multiple add-ons to be loaded together with \feyncalc, \eg \texttt{\$LoadAddOns=\{"AddOn1", "AddOn2", \ldots\}}.

Apart from \feynarts there are also other external packages that can come handy when doing calculations with \feyncalc. With \textsc{FeynRules}~\cite{Christensen:2008py,Alloul:2013bka} it is straightforward to implement new \feynarts models by providing a Lagrangian and deriving the Feynman rules automatically. As far as the analytic evaluation of 1-loop integrals is concerned, the \feynhelpers~\cite{Shtabovenko:2016whf} add-on provides a convenient interface to the corresponding library of \pax.

\subsection{Manual installation}

\feyncalc can be also installed manually by downloading the snapshot directly\footnote{\url{https://github.com/FeynCalc/feyncalc/archive/hotfix-stable.zip}}, unpacking the zip file and copying the \texttt{FeynCalc} directory to \texttt{Applications} inside \texttt{\$User\-Base\-Directory}.\footnote{\texttt{\$UserBaseDirectory} is a global \mma variable that contains the full path to the user base directory. The \texttt{Applications} subdirectory of this directory is the standard place where \mma looks for packages when the user issues \texttt{<<PackageName`}.} To make the documentation work, one should rename the folder \texttt{DocOutput} inside \texttt{FeynCalc} to \texttt{Documentation}. Then, to avoid warnings when opening \feyncalc documentation pages on older \mma versions, it is useful to disable the corresponding warning messages via 
{\small
\begin{mmaCell}[moredefined={FrontEnd}]{Input}
  SetOptions[\$FrontEnd, MessageOptions -> 
  \{"InsufficientVersionWarning" -> False\}]
\end{mmaCell}
}

Many \feyncalc symbols such as 4-vectors, Dirac and color matrices or propagators have extensive typesetting rules attached to them.
However, the output can be typeset only when the default cell output format of the front end is set to \texttt{TraditionalForm} instead of the default \texttt{StandardForm}. It is possible to change this setting only for the current front end session via
{\small
\begin{mmaCell}[moredefined={FrontEndSession,CurrentValue}]{Input}
  CurrentValue[\$FrontEndSession, 
  \{CommonDefaultFormatTypes, "Output"\}] = TraditionalForm
\end{mmaCell}
}
\noindent without affecting the global \mma settings. The same effect can be achieved by adding the line
\begin{verbatim}
$FCTraditionalFormOutput=True;
\end{verbatim}
\noindent
to the special configuration file \texttt{FCConfig.m} inside the \feyncalc directory. The file should be created if missing. 
In general, \texttt{FCConfig.m} may contain arbitrary \mma code that will be evaluated before loading \feyncalc. This can be useful 
to customize the package to the user's requirements.

It is worth noting that \feyncalc also offers two dedicated commands to switch between \texttt{StandardForm} and \texttt{TraditionalForm} output in the new cells on the fly. These are \texttt{FC\-Enable\-Traditional\-FormOutput[]} and \texttt{FC\-Disable\-Traditional\-FormOutput[]}.

\subsection{Development version}

One of our goals is to ensure that when installing \feyncalc the users always obtain a stable version of the package ready for productive use. For this reason every new feature must be carefully tested and validated before it can be added to the stable branch of the repository. The playground where we can freely test different ideas and approaches without worrying about possible side effects is the \texttt{master} branch of the \feyncalc repository, also known as the \textit{development version} of \feyncalc. When this version is considered sufficiently stable and ready for the productive use, its content gets merged into the \texttt{hotfix-stable} branch, which leads to the release of a new stable version of \feyncalc. 

Users that are interested in testing the development version or even employing it on a regular basis (as practiced by \feyncalc enthusiasts) can install the code using the same automatic installer but with an additional option \texttt{InstallFeynCalcDevelopmentVersion} set to \texttt{True}
{\small
\begin{mmaCell}[moredefined={InstallFeynCalc,InstallFeynCalcDevelopmentVersion}]{Input}
  Import@"https://raw.githubusercontent.com/FeynCalc/feyncalc/master/install.m"
  InstallFeynCalc[InstallFeynCalcDevelopmentVersion -> True]
\end{mmaCell}
}
\noindent Any feedback on the development version including inquiries about new features and bug reports is highly encouraged and can be submitted via the official mailing list\footnote{\url{http://www.feyncalc.org/forum/}} or \textsc{GitHub}  Issues.\footnote{\url{https://github.com/FeynCalc/feyncalc/issues}}

\section{New features in \textsc{FeynCalc} 9.3} \label{sec:features}

\subsection{Loading \feyncalc together with other packages}

A common issue that often arises when loading several \mma packages into the same kernel is known as ``shadowing''. 
Suppose that two distinct packages \textsc{A}  and \textsc{B} both define a function \texttt{Foo}. Loading both packages sequentially into the same kernel makes \texttt{Foo} simultaneously appear in multiple contexts (\texttt{A`} and \texttt{B`}), which creates unwanted interference effects between \texttt{A`Foo} and \texttt{B`Foo}. For example, when the user enters \texttt{Foo} in the \mma front end or command line interface, it will depend on the ordering of \texttt{A`} and \texttt{B`} in the global variable \texttt{\$ContextPath}, whether the input will be interpreted as
\texttt{A`Foo} or \texttt{B`Foo}.

In general, when creating new packages it is recommended to include some unique identifier (\eg a shorthand of the package name) to the name of each symbol defined in the package (\eg \texttt{aFoo} and \texttt{bFoo}), so that overlaps between different packages can be avoided. Unfortunately, \mma in no way enforces such naming conventions and many package developers are reluctant to use prefixed symbol names.
Thus, when several packages for similar purposes need to be employed in the same \mma session, there is always a high chance of running into shadowing issues.

\feyncalc and \feynarts constitute a good example of two packages that, for historical reasons, share a large amount of symbols with same names but different definitions. Here our solution was to patch the source code of \feynarts by changing the relevant symbol names and allowing \feyncalc to work only with the modified version of the package. Another approach was taken in~\cite{Shtabovenko:2016whf}, were the relevant packages (\pax or \fire) are deliberately not added to the \texttt{\$ContextPath} and the communication between them and the \feyncalc session is handled using dedicated interfaces. Both workarounds are admittedly not particularly elegant and require permanent maintenance: New versions of the packages may introduce changes that necessitate further adjustments in the patching or interface codes. Given the vast number of existing \mma packages for HEP calculations, one can never hope to make all of them compatible with \feyncalc in a way this was done for \feynarts, \pax or \fire.

As a universal workaround, version 9.3 allows the user to temporarily rename the conflicting symbols of \feyncalc. There are almost no limitations neither on the number of symbols that can be renamed nor on their new names. These changes are not persistent, \ie they do not modify any files on the disk and apply only to the current \feyncalc session. This also implies that all codes evaluated in that session must be compatible to the changes that were introduced when loading the modified \feyncalc. For example, if the user chooses to rename \texttt{Contract} to \texttt{FCContract}, it is mandatory to use the latter name when writing \feyncalc codes in that session, as \texttt{Contract} would have no definition in the \feyncalc context.

The technical implementation of this method relies on the new loading mechanism implemented in \feyncalc 9.3. When the user evaluates \texttt{<<FeynCalc`}, only a very small portion (minimal loader) of the code is directly evaluated on the \mma kernel. The loader will then import the remaining parts of the \feyncalc code as strings and subsequently execute them via \texttt{ToExpression}. At this point it is possible to perform the renaming of almost arbitrary \feyncalc symbols in a very simple way: the code just needs to search for the corresponding name in the strings and replace it accordingly. The list of symbols to be renamed must be specified as a list of replacement rules assigned to the global variable \texttt{\$Rename\-Feyn\-Calc\-Objects}.

For example, suppose that we want to use \feyncalc and \textsc{Lite\-Red}~\cite{Lee:2012cn,Lee:2013mka}, a well-known toolkit for multiloop calculations, on the same kernel. Normally, this leads to the shadowing of \texttt{Factor1}, \texttt{Factor2} and \texttt{MetricTensor}, which happen to be defined in both packages. 
Using the new renaming mechanism one can circumvent this issue by running
{\small
\begin{mmaCell}[index=1,moredefined={FeynCalc,LiteRed,RenameFeynCalcObjects}]{Input}
  \$RenameFeynCalcObjects=\{"Factor1"->"FCFactor1", 
  	"Factor2"->"FCFactor2", "MetricTensor" -> "FCMetricTensor"\};
  <<FeynCalc`
  <<LiteRed`
\end{mmaCell}
}
\noindent Notice that the ordering of the packages does not matter here, in the sense that one can first load \feyncalc and then \textsc{LiteRed} or vice versa. Similarly one can also have \feyncalc loaded together with the reduction package \fire 6~\cite{Smirnov:2019qkx}
{\small
\begin{mmaCell}[moredefined={FeynCalc,FIRE6,RenameFeynCalcObjects}]{Input}
  \$RenameFeynCalcObjects=\{"Contract"->"FCContract"\};
  <<FIRE6`
  <<FeynCalc`
\end{mmaCell}
}
\noindent or the spinor-helicity formalism toolkit \textsc{S@M}~\cite{Maitre:2007jq}
{\small
\begin{mmaCell}[moredefined={FeynCalc,Spinors,SpinorsPath,
	RenameFeynCalcObjects,UserBaseDirectory,FileNameJoin}]{Input}
  \$SpinorsPath = FileNameJoin[\{\$UserBaseDirectory, "Applications", "Spinors"\}];
  \$RenameFeynCalcObjects = \{"Schouten" -> "fcSchouten", "Gamma1" -> "fcGamma1", "Gamma2" -> "fcGamma2", "Gamma3" -> "fcGamma3"\};
  << Spinors`
  << FeynCalc`
\end{mmaCell}
}

While this method might cause some confusion, especially when sharing notebooks with codes that run on top of the ``standard'' or a ``modified'' version of \feyncalc, the resulting benefits vastly outweigh any inconveniences. In particular,  \feyncalc 9.3 can be easily loaded together with any number of other \mma packages without risking any shadowing issues, provided that these packages do not conflict among themselves. This enormously simplifies the task of integrating \feyncalc into complicated computational setups relying on multiple packages and reduces the number of potential pitfalls for new users.

\subsection{UV divergent parts of Passarino--Veltman functions}

One of the most popular applications of \feyncalc is the reduction of 1-loop tensor integrals into linear combinations of Passarino--Veltman functions~\cite{Passarino:1978jh}, which can be then evaluated using publicly available codes such as \pax, \textsc{LoopTools}~\cite{Hahn:1998yk}, \textsc{Collier}~\cite{Denner:2016kdg,Denner:2002ii,Denner:2005nn,Denner:2010tr}, \textsc{QCDLoop}~\cite{Ellis:2007qk}, \textsc{OneLOop}~\cite{vanHameren:2010cp} etc.

Sometimes one is interested only in the ultraviolet (UV) divergent parts of the occurring Passarino--Veltman functions, \eg when calculating renormalization constants using minimal subtraction or performing a quick check for the cancellation of the UV poles
in a renormalized amplitude. For functions with up to 4 legs this functionality is already present in \pax and can be accessed via the \feynhelpers routine \texttt{PaX\-Evaluate\-UV\-Part}. \feyncalc 9.3 offers an alternative implementation of this capability that does not rely on \pax and works with an arbitrary number of external legs. The new routine \texttt{PaVeUVPart} is based on the algorithm\footnote{The \mma notebook with the code is contained in the source of the arXiv preprint for~\cite{Sulyok:2006xp}. It is self-contained and does not require any external packages.} developed by Georg Sulyok~\cite{Sulyok:2006xp}, who kindly permitted us to integrate his code into \feyncalc. 

Since \texttt{PaVeUVPart} operates on \texttt{PaVe} functions, all 1-loop integrals in the input must be first converted 
to this notation either by using \texttt{ToPaVe} (for scalar integrals) or by employing \texttt{TID} with the options \texttt{UsePaVeBasis} and \texttt{ToPaVe} both set to \texttt{True} (for tensor integrals). For example, we can readily obtain the UV divergence of the tadpole $A_0(m^2)$ by issuing
{\small
\begin{mmaCell}[index=1,moredefined={PaVeUVPart,PaVe}]{Input}
  PaVeUVPart[PaVe[0, \{\}, \{m^2\}]]
\end{mmaCell}

\begin{mmaCell}{Output}
  -\mmaFrac{2\,\mmaSup{m}{2}}{D-4}
\end{mmaCell}
}
\noindent
A less trivial application is the analysis of the fully massive 4-point coefficient function $D_{000000}$ with full kinematic dependence, which immediately returns the correct answer (\cf \eg~\cite{Denner:2005nn})
{\small
\begin{mmaCell}[index=2,moredefined={PaVeUVPart,PaVe}]{Input}
  PaVeUVPart[PaVe[0,0,0,0,0,0, \{p11, p12, p23, p33, p22, p13\}, 
  \{m0^2, m1^2, m2^2, m3^2\}]]
\end{mmaCell}

\begin{mmaCell}[index=3]{Output}
  \mmaFrac{-5\,\mmaSup{m0}{2}-5\,\mmaSup{m1}{2}-5\,\mmaSup{m2}{2}-5\,\mmaSup{m3}{2}+p11+p12+p13+p22+p23+p33}{480(D-4)}
\end{mmaCell}
}

It is important to keep in mind that Passarino--Veltman functions may exhibit not only UV but also infrared (IR) divergences. Poles in $1/\varepsilon$ due to soft or collinear singularities\footnote{
The double poles $1/\varepsilon^2$ caused by simultaneous soft and collinear singularities are also possible.} will mix with $1/\varepsilon$ from UV divergences, unless special care was taken to employ different regulators for UV and IR. This is also the reason why the scaleless 2-point function $B_0(0,0,0)$ can be put to zero, despite being proportional to $(1/\varepsilon_{\textrm{UV}} - 1/\varepsilon_{\textrm{IR}})$. Since \texttt{PaVeUVPart} only deals with UV singularities, it cannot tell anything about possible IR  divergences of the occurring Passarino--Veltman functions. For example, a $C_0$ function is always UV finite, but may exhibit IR poles depending on the kinematic invariants in its arguments. Therefore, here \texttt{PaVeUVPart} will always return $0$ (which is correct), but the UV finiteness in no way implies that $C_0$ is also IR finite.

In order to correctly extract UV divergences of log divergent scaleless loop integrals (such as $B_0(0,0,0)$ or $C_{00}(0,0,0,0,0,0)$), that would be otherwise discarded as scaleless, one should use the global switch \texttt{\$Keep\-Log\-Divergent\-Scaleless\-Integrals}. This forces \feyncalc not to put such integrals to zero and allows the package to consistently disentangle UV and IR divergences at 1-loop, provided that no Integration-By-Parts~\cite{Chetyrkin:1981qh,Tkachov:1981wb} reductions are employed\footnote{Existing IBP codes always mix UV and IR divergences, thus making it impossible to trace back their origin.}. For example, one can consider the scaleless tensor integral $\int d^D l \, l^\mu l^\nu / l^6$ and convince oneself that it indeed has a UV pole, that is normally canceled by an IR pole with an opposite sign, making the whole integral vanish
{\small
\begin{mmaCell}[index=3,moredefined={PaVeUVPart,FVD,FAD,TID, ToPaVe,KeepLogDivergentScalelessIntegrals}]{Input}
  \$KeepLogDivergentScalelessIntegrals = True;
  TID[FVD[l, \(\mu\)] FVD[l, \(\nu\)] FAD[\{l, 0, 3\}], l, ToPaVe -> True] // PaVeUVPart
\end{mmaCell}
{\small
\begin{mmaCell}{Output}
  -\mmaFrac{2\,i\,\mmaSup{\(\pi\)}{2}\mmaSup{g}{\(\mu\nu\)}}{(D-4)\,D}
\end{mmaCell}
}
}
\noindent The ability to investigate such subtleties with very little effort makes \texttt{PaVe\-UV\-Part} (especially in conjunction with \texttt{\$Keep\-Log\-Divergent\-Scaleless\-Integrals}) a very useful addition to the \feyncalc toolkit.

\subsection{Majorana spinors}
Majorana fermions appear in the spectrum of numerous HEP models, ranging from Standard Model extensions to special theories that help to improve our understanding of QCD. A prominent representative of the latter category is $\mathcal{N}=4$ Supersymmetric Yang--Mills (SYM) ~\cite{Brink:1976bc,Gliozzi:1976qd}, where for suitable observables the principle of maximal transcendentality~\cite{Kotikov:2001sc,Kotikov:2002ab,Kotikov:2004er} is conjectured to link the highest transcendental weight pieces of pQCD results to those in $\mathcal{N}=4$ SYM. Even in cases when the principle is not satisfied exactly, the $\mathcal{N}=4$ SYM and pQCD results may still share the basis of master integrals and be composed of the same set of building block functions. This is why for computationally challenging observables a simpler calculation in $\mathcal{N}=4$ SYM
can be often regarded as the first step towards the full pQCD result.
One of the recent examples for the application of this strategy is the analytic fixed-order result for the Energy--Energy correlations~\cite{Basham:1978bw} in $e^+ e^-$ annihilation . The calculation of this event shape variable at NLO in $\mathcal{N}=4$ SYM was presented in~\cite{Belitsky:2013ofa} and turned out to be very helpful in the task of tackling the more challenging pQCD case~\cite{Dixon:2018qgp,Gituliar:2017umx}. This story could repeat itself also at NNLO, where the $\mathcal{N}=4$ SYM result was made available recently~\cite{Henn:2019gkr}.

From the technical point of view, evaluation of Feynman diagrams with Majorana particles involves several subtleties that do not arise in calculations with Dirac fermions. First of all, depending on the chosen prescription to handle such particles, one may end up with Feynman rules that contain explicit charge-conjugation matrices and transposed Dirac matrices, which introduces an additional layer of complexity. Second, the determination of the relative sign between diagrams is much less straightforward as in the Dirac case due to the possible violations of the fermion number flow.

A simple and unambiguous set of prescriptions for writing down amplitudes with Majorana spinors that circumvents the above-mentioned issues was presented in~\cite{Denner:1992vza,Denner:1992me}\footnote{\cf also~\cite{Paraskevas:2018mks} for recent developments in this direction.} and subsequently implemented in \feynarts. This implies that the amplitudes with Majorana spinors generated by \feynarts contain only the usual Dirac matrices $\gamma^\mu$ and $\gamma^5$ as well as the four-spinors $u$, $v$, $\bar{u}$ and $\bar{v}$, while the relative signs between diagrams have already been fixed.\footnote{
In the presence of vertices that contain more than two fermion fields the amplitudes need to be post-processed to build correct fermion chains and fix the signs.} Furthermore, the output is free of transposed Dirac matrices and explicit charge-conjugation matrices.

However, even under these favorable conditions, special care is still required when squaring amplitudes with Majorana spinors. The simplest approach to calculate unpolarized matrix elements squared $|\mathcal{M}|^2$ in diagrams with spinor particles is to apply the standard spin sum formulas
\begin{equation}
\sum_s u(p,s) \bar{u}(p,s) = \slashed{p}+m, \quad \sum_s v(p,s) \bar{v}(p,s) = \slashed{p}-m, \label{eq:spinsums}
\end{equation}
thus converting products of closed spinor chains into Dirac traces. Unfortunately, in the presence of Majorana fermions the contributions from interference diagrams may involve terms which do not allow for a direct application of the formulas from \Eq\eqref{eq:spinsums}. This issue can be resolved by transposing some of the spinor chains and introducing the charge conjugation matrix $C$ with (\cf \eg ~\cite{Pal:2010ih} for a pedagogical treatment)
\begin{align}
v^T(p,s) &= - \bar{u} (p,s) C, \quad \bar{u}^T(p,s) = C^{-1} v(p,s), \nonumber \\
C \gamma^{\mu T} C^{-1} &= - \gamma^\mu, \quad C \gamma^{5 T} C^{-1} = \gamma^5.
\end{align}

In \feyncalc 9.3 the default function for rewriting products of closed spinor chains as Dirac traces, \texttt{FermionSpinSum}, can automatically reorder the chains when necessary, as in the following example where formulas from \Eq\eqref{eq:spinsums} cannot be applied straightforwardly
\begin{equation}
\bar{u}(k_2) P_R v(k_1) \,\, \bar{v}(p_1) P_R u(p_2) \,\, \bar{u}(p_1) P_L u(k_2)  \,\,  \bar{u}(p_2) P_L u(k_1) \label{eq:spinprod}
\end{equation}
{\small
\begin{mmaCell}[moredefined={FermionSpinSum,SpinorU,SpinorV,
SpinorUBar,SpinorVBar,GA}]{Input}
  FermionSpinSum[SpinorUBar[k2, m2].GA[6].SpinorV[k1, m1] * SpinorVBar[p1, m1].GA[6].SpinorU[p2, m2] * SpinorUBar[p1, m1].GA[7].SpinorU[k2, m2] * SpinorUBar[p2, m2].GA[7].SpinorU[k1, m1]]
\end{mmaCell}
\begin{mmaCell}{Output}
  -tr((\mmaOver{\(\gamma\)}{_}\(\cdot\)\mmaOver{k1}{_}+m1).\mmaSup{\mmaOver{\(\gamma\)}{_}}{6}.(m2-\mmaOver{\(\gamma\)}{_}\(\cdot\)\mmaOver{k2}{_}).\mmaSup{\mmaOver{\(\gamma\)}{_}}{7}.(\mmaOver{\(\gamma\)}{_}\(\cdot\)\mmaOver{p1}{_}-m1).\mmaSup{\mmaOver{\(\gamma\)}{_}}{6}.(\mmaOver{\(\gamma\)}{_}\(\cdot\)\mmaOver{p2}{_}+m2).\mmaSup{\mmaOver{\(\gamma\)}{_}}{7})
\end{mmaCell}
}
\noindent The automatic reordering can be turned off by setting the option \texttt{Spinor\-Chain\-Transpose} to \texttt{False}. This can be used as a consistency check, since amplitudes involving only Dirac spinors should not require any reordering. 

Furthermore, we added a dedicated function \texttt{Spinor\-Chain\-Transpose} that transposes closed spinor chains as in
{\small
\begin{mmaCell}[moredefined={SpinorChainTranspose,SpinorU,SpinorV,
SpinorUBar,SpinorVBar,GA}]{Input}
  SpinorChainTranspose[SpinorVBar[p1, m1].GA[6].SpinorU[p2, m2]]
\end{mmaCell}
\begin{mmaCell}{Output}
  -(\(\varphi\)(-\mmaOver{p2}{_},m2)).\mmaSup{\mmaOver{\(\gamma\)}{_}}{6}.(\(\varphi\)(\mmaOver{p1}{_},m1))
\end{mmaCell}
}
\noindent Since it is normally not useful to transpose every chain in the expression, the value of the option \texttt{Select} specifies types of the chains that should be processed. By default, these are all chains that start with a $\bar{v}$-spinor and end with a $u$- or $v$-spinor, where the spinors may depend on 4 or $D$-dimensional momenta.
{ \small
\begin{mmaCell}[moredefined={SpinorChainTranspose,OptionValue,Select,
StandardForm}]{Input}
  OptionValue[SpinorChainTranspose, Select] // StandardForm
\end{mmaCell}

\begin{mmaCell}{Output}
  \{\{SpinorVBar[_,_],SpinorU[_,_]\},
   \{SpinorVBar[_,_],SpinorV[_,_]\},
   \{SpinorVBarD[_,_],SpinorUD[_,_]\},
   \{SpinorVBarD[_,_],SpinorVD[_,_]\}\}
\end{mmaCell}
}
\noindent Using patterns one can also provide more sophisticated selection criteria.

Finally, the operation of transposing a chain of Dirac matrices $X$ and sandwiching it between $C$ and $C^{-1}$ is provided through the new function \texttt{FCCharge\-Conjugate\-Transposed} (abbreviated with \texttt{FCCCT}). By default, the expression remains in the unevaluated form as $C X^T C^{-1}$
{\small
\begin{mmaCell}[moredefined={FCCCT,SpinorU,SpinorV,SpinorUBar,SpinorVBar,GA,GS}]{Input}
  FCCCT[GA[7].(GS[k2] + m2).GA[6]]
\end{mmaCell}
\begin{mmaCell}{Output}
  C\mmaSup{(\mmaSup{\mmaOver{\(\gamma\)}{_}}{7}.(\mmaOver{\(\gamma\)}{_}\(\cdot\)\mmaOver{k2}{_}+m2).\mmaSup{\mmaOver{\(\gamma\)}{_}}{6})}{T}\mmaSup{C}{-1}
\end{mmaCell}
}
\noindent The explicit result can be obtained by setting the option \texttt{Explicit} to \texttt{True} or by applying the function \texttt{Explicit} to the expression
{\small
\begin{mmaCell}[moredefined={FCCCT,SpinorU,SpinorV, SpinorUBar,SpinorVBar,GA,GS,Explicit}]{Input}
  FCCCT[GA[7].(GS[k2] + m2).GA[6]]//Explicit
\end{mmaCell}
\begin{mmaCell}{Output}
  \mmaSup{\mmaOver{\(\gamma\)}{_}}{6}.(m2-\mmaOver{\(\gamma\)}{_}\(\cdot\)\mmaOver{k2}{_}).\mmaSup{\mmaOver{\(\gamma\)}{_}}{7}
\end{mmaCell}
}
\noindent Hence, while \feyncalc 9.2 would fail to construct the trace for an expression similar to that in \Eq\eqref{eq:spinprod}, the new version 9.3 not only bypasses these limitations in a transparent way, but also allows the user to explicitly reorder selected spinor chains or to apply the charge conjugation operation to arbitrary strings of Dirac matrices.

\subsection{Dirac matrices and spinors with explicit indices}
The internal \feyncalc representation of Dirac algebra is built upon the symbols \texttt{DiracGamma} and \texttt{Spinor}. Both are defined as noncommutative objects and their products must be written using \texttt{Dot} (``.'', noncommutative multiplication operator) instead of \texttt{Times} (``*'', commutative multiplication operator). The Dirac indices are always suppressed, which is sufficient as long as one works only with closed spinor chains and Dirac traces. A single open chain  (\eg $\left (\gamma^\mu (\slashed{p}+m) \gamma^\nu \right )_{ij}$ written as \texttt{GAD[$\mu$].(GSD[p]+$m$).GAD[$\nu$]}) is still unambiguous even without explicit Dirac indices. However, once the number of the uncontracted Dirac indices exceeds 2 (\eg in $\gamma^{\mu}_{ij}\gamma^\nu_{kl}$) ambiguities become unavoidable.

In order to mitigate these shortcomings \feyncalc 9.3 features an initial support for explicit Dirac indices via special heads \texttt{DiracIndex}, \texttt{DiracIndex\-Delta} and \texttt{Dirac\-Chain}. Then, $\gamma^{\mu}_{ij}\gamma^\nu_{kl}$ can be directly written as
{\small
\begin{mmaCell}[moredefined={DCHN,GA}]{Input}
  DCHN[GA[\(\mu\)], i, j] DCHN[GA[\(\nu\)], k, l]
\end{mmaCell}
\begin{mmaCell}{Output}
  \mmaSub{(\mmaSup{\mmaOver{\(\gamma\)}{_}}{\(\mu\)})}{ij}\mmaSub{(\mmaSup{\mmaOver{\(\gamma\)}{_}}{\(\nu\)})}{kl}
\end{mmaCell}
}
\noindent where we used the built-in shortcut \texttt{DCHN} for \texttt{Dirac\-Chain}. Internally, \feyncalc understands
\texttt{DCHN[GA[$\mu$], i, j]} as 
\begin{mmaCell}[addtoindex=2,form=StandardForm]{Output}
  DiracChain[DiracGamma[LorentzIndex[\(\mu\)]],DiracIndex[i],DiracIndex[j]]
\end{mmaCell}
The last two arguments of a three argument \texttt{Dirac\-Chain} may also contain spinors as in
{\small
\begin{mmaCell}[moredefined={DCHN,GA,GS,GSD,SpinorV,SpinorUBar}]{Input}
  \{DCHN[GS[q], SpinorUBar[p1], i], DCHN[GSD[q], j, SpinorV[p2]], DCHN[GS[q], SpinorUBar[p1], SpinorV[p2]]\}
\end{mmaCell}
\begin{mmaCell}[moredefined={DCHN,GA,GS,SpinorV,SpinorUBar}]{Output}
  \{\mmaSub{(\mmaOver{u}{_}(p1).\mmaOver{\(\gamma\)}{_}\(\cdot\)\mmaOver{q}{_})}{i},\mmaSub{(\(\gamma\cdot\)q.v(p2))}{j},(\mmaOver{u}{_}(p1).\mmaOver{\(\gamma\)}{_}\(\cdot\)\mmaOver{q}{_}.v(p2))\}
\end{mmaCell}
}
\noindent while a \texttt{Dirac\-Chain} with just two arguments is used to represent spinor bilinears or standalone spinors
{\small
\begin{mmaCell}[moredefined={DCHN,GA,GS,SpinorV,SpinorUBar}]{Input}
  \{DCHN[SpinorUBar[p1], i], DCHN[j, SpinorV[p2]], 
  DCHN[SpinorUBar[p1], SpinorV[p2]]\}
\end{mmaCell}

\begin{mmaCell}[]{Output}
  \{\mmaSub{(\mmaOver{u}{_}(p1))}{i},\mmaSub{(v(p2))}{j},(\mmaOver{u}{_}(p1).v(p2))\}
\end{mmaCell}
}
\noindent As far as the Kronecker delta in the Dirac space $\delta_{ij}$ is concerned, it must be entered as \texttt{DIDelta[i, j]}.
Of course, several routines have been added to manipulate the new indexed objects in a convenient way. \texttt{DiracChainJoin} can be used to contract dummy indices
{\small
\begin{mmaCell}[moredefined={DCHN,GA,GS,SpinorV,SpinorUBar,DIDelta,DiracChainJoin}]{Input}
  DCHN[GA[\(\mu\)], i, j] DIDelta[j, k] DCHN[GA[\(\nu\)], k, l]
  DiracChainJoin[\%]
\end{mmaCell}

\begin{mmaCell}[]{Output}
  \mmaSub{\(\delta\)}{jk}\mmaSub{(\mmaSup{\mmaOver{\(\gamma\)}{_}}{\(\mu\)})}{ij}\mmaSub{(\mmaSup{\mmaOver{\(\gamma\)}{_}}{\(\nu\)})}{kl}
\end{mmaCell}

\begin{mmaCell}{Output}
  \mmaSub{(\mmaSup{\mmaOver{\(\gamma\)}{_}}{\(\mu\)}.\mmaSup{\mmaOver{\(\gamma\)}{_}}{\(\nu\)})}{il}
\end{mmaCell}
}
\noindent while \texttt{DiracChainFactor} is useful to pull non-Dirac symbols out of an indexed Dirac chain
{\small
\begin{mmaCell}[moredefined={DCHN,GA,GS,SpinorV,SpinorUBar,DIDelta,DiracChainFactor}]{Input}
  DCHN[2 x GS[q], i, j]
  DiracChainFactor[\%]
\end{mmaCell}

\begin{mmaCell}[]{Output}
  \mmaSub{(2\,x\,\mmaOver{\(\gamma\)}{_}\(\cdot\)\mmaOver{q}{_})}{ij}
\end{mmaCell}

\begin{mmaCell}{Output}
  2\,x\,\mmaSub{(\mmaOver{\(\gamma\)}{_}\(\cdot\)\mmaOver{q}{_})}{ij}
\end{mmaCell}
}
\noindent To expand sums inside an indexed Dirac chain one can employ \texttt{DiracChainExpand}
{\small
\begin{mmaCell}[moredefined={DCHN,GA,GS,SpinorV,SpinorUBar,DIDelta,DiracChainExpand}]{Input}
  DCHN[GS[q] + m, i, j]
  DiracChainExpand[\%]
\end{mmaCell}

\begin{mmaCell}[]{Output}
  \mmaSub{(\mmaOver{\(\gamma\)}{_}\(\cdot\)\mmaOver{q}{_}+m)}{ij}
\end{mmaCell}

\begin{mmaCell}{Output}
  \mmaSub{(\mmaOver{\(\gamma\)}{_}\(\cdot\)\mmaOver{q}{_})}{ij}+m \mmaSub{(1)}{ij}
\end{mmaCell}
}
\noindent with \texttt{DiracChainCombine} being nearly the inverse of this operation
{\small
\begin{mmaCell}[moredefined={DCHN,GA,GS,SpinorV,SpinorUBar,DIDelta,DiracChainCombine}]{Input}
  DCHN[GA[\(\mu\)].GS[q], i, j] + m DCHN[GA[\(\mu\)], i, j]
  DiracChainCombine[\%]
\end{mmaCell}

\begin{mmaCell}[]{Output}
  m \mmaSub{(\mmaSup{\mmaOver{\(\gamma\)}{_}}{\(\mu\)})}{ij}+\mmaSub{(\mmaSup{\mmaOver{\(\gamma\)}{_}}{\(\mu\)}.(\mmaOver{\(\gamma\)}{_}\(\cdot\)\mmaOver{q}{_}))}{ij}
\end{mmaCell}

\begin{mmaCell}{Output}
  \mmaSub{(m \mmaSup{\mmaOver{\(\gamma\)}{_}}{\(\mu\)}+\mmaSup{\mmaOver{\(\gamma\)}{_}}{\(\mu\)}.(\mmaOver{\(\gamma\)}{_}\(\cdot\)\mmaOver{q}{_}))}{ij}
\end{mmaCell}
}
Additional functions for working with indexed Dirac chains are planned for the future. Still, already the built-in
functionality turns out to be surprisingly powerful when approaching tasks that require manipulations of explicit
Dirac indices. For example, \feyncalc 9.3 is now able to determine relative signs of amplitudes with 4-fermion vertices
generated from suitable \feynarts models. The related algorithm was adopted from \formcalc and implemented in an auxiliary routine \texttt{FCFADiracChainJoin}.\footnote{One of the authors (V.\,S.) would like to acknowledge 
David Straub for bringing his attention to this issue and the suggestion to determine relative signs in the same way as it is done in \formcalc. Thomas Hahn is acknowledged for helpful explanations regarding the relevant \formcalc code.}

When converting \feynarts output into a proper \feyncalc expression, 
\texttt{FCFAConvert} will automatically call \texttt{FC\-FA\-Dirac\-Chain\-Join}, closing the chains accordingly
and obtaining correct relative signs, so that no additional user interaction is required. As usual, this behavior can be disabled by setting the \texttt{FC\-FA\-Dirac\-Chain\-Join} option of 
\texttt{FCFAConvert} to \texttt{False}. Needless to say that this level of convenience became possible only after support for indexed Dirac chains became part of \feyncalc.

\section{Efficient usage of \feyncalc in loop calculations} \label{sec:usage}

In this section we would like to address some common mistakes, misconceptions and pitfalls
that often affect \feyncalc users  and make it more challenging for them to use the package in real-life calculations. The topics have been chosen from our subjective experience of answering some recurring questions via the official mailing list or in private communication.

\subsection{Faster 1-loop tensor reduction}

Tensor reduction of 1-loop integrals in \feyncalc is handled by \texttt{TID}. Unlike the old and currently deprecated
\texttt{OneLoop}, the default working mode of \texttt{TID} is to reduce each integral to
the scalar Passarino--Veltman functions $A_0$ (tadpole), $B_0$ (bubble), $C_0$ (triangle), $D_0$ (box) etc. On the one hand, this method
is often desirable for analytic studies since it leads to a minimal number of master
integrals. On the other hand, doing full reduction of tensor integrals that depend on many kinematic invariants inevitably generates page-long expressions with master integrals being multiplied by very large and complicated prefactors. 

Contrary to that, the algorithm implemented in \texttt{OneLoop} performs only a partial reduction, which expresses the result in terms of Passarino--Veltman coefficient functions (\eg $A_{00}$, $B_1$, $C_{12}$ etc.). The coefficient functions can be reduced to the scalar functions by means of \texttt{PaVeReduce}, so that they do not represent the most compact basis of 1-loop integrals. However, their prefactors are just tensors made out of 4-momenta and the metric. There are no huge polynomials of kinematic invariants and the results written in terms of these functions appear very compact. Furthermore, the so-obtained results are
equally suitable for numerical and symbolic evaluation. 

 It is obvious that the full reduction \`a la \texttt{TID} requires significantly more time than the partial one \`a la \texttt{OneLoop}, especially if the kinematics is completely generic.\footnote{For performance reasons, it is crucial to specify as much kinematic invariants as possible, before performing the reduction. Doing it the other way around can be orders of magnitude slower.} As a consequence, we have often received complaints that while the old \texttt{OneLoop} was very fast, \texttt{TID} is very slow. The simple solution to this problem, is to change the working mode of \texttt{TID}, so that it will also perform a partial reduction into coefficient functions. The relevant options is called \texttt{UsePaVeBasis}. Setting it to \texttt{True} as in
 
\begin{mmaCell}[moredefined={TID,UsePaVeBasis,ToPaVe,amp}]{Input}
  TID[amp, q, UsePaVeBasis->True]
\end{mmaCell}
leads to a very fast tensor reduction and the output is similar to what one would have obtained with \texttt{OneLoop}.

\subsection{Treatment of \texorpdfstring{$\gamma^5$}{gamma5} in dimensional regularization}

The precise treatment of $\gamma^5$ in dimensional regularization within \feyncalc is a recurring question among \feyncalc users. It is a well-known fact (\cf \eg~\cite{Jegerlehner:2000dz}) that the definition of $\gamma^5$ in 4 dimensions cannot be consistently extended to $D$ dimensions without giving up either the anticommutativity property $\{\gamma^5, \gamma^\mu\} = 0$ or the cyclicity of the Dirac trace.

Over the years people developed numerous prescriptions (schemes) to circumvent these issues
and arrive at physical results. Nonetheless, as of today there seems to be no universally
accepted solution that can be readily applied to any model at any loop order in a fully automatic fashion. This is mainly related to the fact that calculations with $\gamma^5$ are not limited to the algebraic manipulations of Dirac matrices. In addition to that one is usually required to check that the final result preserves all the relevant symmetries, \eg generalized Ward identities and Bose symmetry. If some of those symmetries are violated due to the chosen $\gamma^5$ scheme, they must be restored by hand, \eg by introducing finite counterterms. Unfortunately, an explicit determination of such counterterms for a given model is a nontrivial task, especially beyond 1-loop.\footnote{Interesting comments and further references can be found \eg in Chapter D of~\cite{Blondel:2018mad}. For SM calculations \cf~\cite{Trueman:1995ca,Denner:2019vbn}.}

This is why we would like to make clear that this nonalgebraic part of a $\gamma^5$-calculation is not the task of \feyncalc but the duty of the user. One should not naively expect that a mere tool for algebraic calculations such as \feyncalc can replace physical insight into the problem at hand, especially with such a subtle issue as $\gamma^5$. Therefore, the answer to the question whether \feyncalc could automatically handle $\gamma^5$ in such a way, that the user can simply read off the final result without any further thoughts about $\gamma^5$ is a clear ``no''.

The responsibility of \feyncalc is to ensure that algebraic manipulations of Dirac matrices
(including $\gamma^5$) are consistent within the chosen scheme. For that purpose \feyncalc implements two\footnote{It is also possible to use Larin's scheme~\cite{Larin:1993tq} but the underlying code is still poorly optimized and hence rather slow. A complete rewrite is scheduled for the future.} ways to handle $\gamma^5$ in $D$-dimensions.

The Naive or Conventional Dimensional Regularization (NDR or CDR respectively)~\cite{Chanowitz:1979zu} \textit{assumes} that one can define a $D$-dimensional $\gamma^5$ that anticommutes with any other Dirac matrix and does not break the cyclicity of the trace. For \feyncalc this means that in every string of Dirac matrices all $\gamma^5$ can be safely anticommuted to the right end of the string. In the course of this operation \feyncalc can always apply $(\gamma^5)^2 = 1$.
Consequently, all Dirac traces with an even number of $\gamma^5$ can be rewritten as traces that involve only the first four $\gamma$-matrices and evaluated directly. The problematic cases are $\gamma^5$-odd traces with an even number of other Dirac matrices, where the $\mathcal{O}(D-4)$ pieces of the result depend on the initial position of $\gamma^5$ in the string. Using the anticommutativity property they can be always rewritten as traces of a string of other Dirac matrices and one $\gamma^5$. If the number of the other Dirac matrices is odd, such a trace is put to zero \ie
\begin{equation}
\Tr(\gamma^{\mu_1} \ldots \gamma^{\mu_{2n-1}} \gamma^5) = 0, \quad n \in \mathbb{N}
\end{equation}
If the number is even, the trace
\begin{equation}
\Tr(\gamma^{\mu_1} \ldots \gamma^{\mu_{2n}} \gamma^5)
\end{equation}
is returned unevaluated, since \feyncalc does not know how to calculate it in a consistent way. A user who knows how these ambiguous objects should be treated in the particular calculation can still take care of the remaining traces by hand. This ensures that the output produced by \feyncalc is algebraically consistent to the maximal extent possible
in the NDR scheme without extra assumptions.

The other $\gamma^5$ scheme available in the package is the Breitenlohner--Maison implementation~\cite{Breitenlohner:1977hr} of the t'Hooft--Veltman~\cite{tHooft:1972tcz} prescription, often abbreviated as BMHV, HVBM, HV or BM scheme. In this approach $\gamma^5$ is treated as a purely 4-dimensional object, while $D$-dimensional Dirac matrices and 4-vectors are decomposed into $4$- and $D-4$-dimensional components. Following~\cite{Buras:1989xd} \feyncalc typesets the former with a bar and the latter with a hat \eg
\begin{equation}
\gamma^\mu = \bar{\gamma}^\mu + \hat{\gamma}^\mu, \quad p^\mu = \bar{p}^\mu + \hat{p}^\mu
\end{equation}
The main advantage of the BMHV scheme is that the Dirac algebra (including traces) can be evaluated without any algebraic ambiguities. However, calculations involving tensors from three different spaces ($D$, $4$ and $D-4$) often turn out to be rather cumbersome, even when using computer codes. Moreover, this prescription is known to artificially violate Ward identities in chiral theories, which is something that can be mostly avoided when using NDR. Within BMHV \feyncalc can simplify arbitrary strings of Dirac matrices and calculate arbitrary traces out-of-the-box. The evaluation of $\gamma^5$-odd Dirac traces is performed using the  West-formula from~\cite{West:1991xv}. It is worth noting that $D-4$-dimensional components of external momenta are not set to zero by default, as it is conventionally done in the literature.
If this is required, the user should evaluate \texttt{Momentum[$p_i$,D-4]=0} for each relevant momentum $p_i$. To remove such assignments one should use \texttt{FC\-Clear\-Scalar\-Products[]}.

In \feyncalc 9.3 the mechanism for switching between different schemes was reworked to be more consistent and user friendly. The new standard way to select a scheme is to use the function
\texttt{FCSetDiracGammaScheme[]}.\footnote{The usage of old switches \texttt{\$BreitMaison} and \texttt{\$Larin} is still supported for compatibility reasons but is considered deprecated.
The \texttt{\$West} switch was removed, but the same effect can be achieved by setting the option \texttt{West} of \texttt{DiracTrace} to \texttt{True} (default) or \texttt{False}.} Currently, it supports following arguments: \texttt{"NDR"}, \texttt{"NDR-Discard"}, \texttt{"BMHV"} and \texttt{"Larin"}. Here \texttt{"NDR-Discard"} constitutes a special variety of the \texttt{"NDR"} scheme where the remaining $\gamma^5$-odd traces are set to zero. The name of the current scheme can be requested using \texttt{FC\-Get\-Dirac\-Gamma\-Scheme[]}. NDR still remains the default scheme used in \feyncalc.

\subsection{Taking the correct limit \texorpdfstring{$D\to4$}{D to 4}}

When doing calculations in dimensional regularization it is often necessary to promote 4-dimensional quantities (\eg scalar products or Dirac matrices) to $D$-dimensional ones and vice versa. Another task that often arises in this context is the expansion of the obtained results around $D=4$ in the regularization parameter $\varepsilon$. Although \feyncalc is equipped with special functions for automatizing these operations, sometimes the users do not succeed in employing them correctly. Therefore, it appears necessary to reiterate the basics of handling objects living in different dimensions when using the package.

In order to change the dimension of Dirac matrices, 4-vectors, scalar products and other suitable quantities, 
one should use \texttt{Change\-Dimension}, where the first argument is the input expression and the second one is the dimension. It is important to stress that \texttt{Change\-Dimension} does not operate on standalone $D$-symbols, which often appear in the prefactors of Dirac spinor chains or loop integrals. This can be seen \eg when applying the function to a $D$-dimensional integral with a $D$-dependent prefactor and inspecting the result
{\small
\begin{mmaCell}[moredefined={ChangeDimension,FAD,FVD,FCE,StandardForm}]{Input}
  ChangeDimension[(D-3)/(D-2) FAD[l, l+p] FVD[l, \(\mu\)] FVD[l, \(\nu\)], 4] // FCE // StandardForm
\end{mmaCell}
\begin{mmaCell}{Output}
  \mmaFrac{(-3+D) FAD[l,l+p,Dimension\(\to\)4] FV[l,\(\mu\)] FV[l,\(\nu\)]}{-2+D}
\end{mmaCell}
}
\noindent where we see that the prefactor has not been changed. This behavior is intended, since 
\feyncalc contains a dedicated function that covers the other case:  \texttt{FCReplaceD} replaces all occurrences of $D$ in the prefactors without altering the dimensions of matrices and tensors. Applying \texttt{FCReplaceD} to the expression
from the previous example we can immediately observe the difference to the action of \texttt{ChangeDimension}
{\small
\begin{mmaCell}[moredefined={FCReplaceD,FAD,FVD,FCE,StandardForm,Epsilon}]{Input}
  FCReplaceD[(D-3)/(D-2)  FAD[l, l+p] FVD[l,\(\mu\)] FVD[l, \(\nu\)], D->4-2 Epsilon] // FCE // StandardForm
\end{mmaCell}
\begin{mmaCell}{Output}
  \mmaFrac{(1-2 Epsilon) FAD[l,l+p] FVD[l,\(\mu\)] FVD[l,\(\nu\)]}{2-2 Epsilon}
\end{mmaCell}
}
\noindent The biggest advantage of using \texttt{ChangeDimension} and \texttt{FCReplaceD} as compared to naive replacement rules, is that only the former method allows one to distinguish between prefactors, matrices and tensors. In most cases doing something similar to \texttt{exp\,/.\,D->4} or \texttt{exp\,/.\,D->4-2 Epsilon} will not produce the desired result.

In fact, the naive application of the \texttt{D->4} rule to dimensionally regularized  expressions appears to be one of the most common mistakes that we observe in the practice of using \feyncalc in loop calculations. One should always keep in mind that the product of the $\mathcal{O}(\varepsilon)$ piece of the prefactor and the $1/\varepsilon$-pole of a divergent loop integral generates a finite $\mathcal{O}(\varepsilon^0)$ contribution to the final result which cannot be discarded. Schematically, at 1-loop one may often observe terms that look like
\begin{equation}
f(D) \mathcal{I}_D = \left ( f_0 + f_1 \varepsilon + \ldots \right ) \left ( \frac{c_{-1}}{\varepsilon} + c_0 + \ldots \right ).
\end{equation}
Replacing $f(D)$ with $f(4)$ in this expression is equivalent to dropping the contribution from $f_1 c_{-1}$, while keeping the other finite piece $f_0 c_0$, which is obviously inconsistent. Therefore, naive simplifications of prefactors by setting $D=4$ instead of expanding the full expression \textit{around} $D=4$ will most likely yield wrong results and should be avoided.

\subsection{Debugging \feyncalc functions}

When a \feyncalc function behaves in a strange way by generating error messages, aborting the evaluation or simply requiring too much time to finish, one might be interested to understand the cause for this behavior without contacting the developers.\footnote{Of course, the developers always appreciate bug reports and interesting examples of performance bottlenecks.} This can be achieved by activating the debugging output, which is available for most high-level functions (\eg \texttt{TID}, \texttt{DiracSimplify}, \texttt{Contract} etc.) via the special option \texttt{FCVerbose}. For example, evaluating
{\small
\begin{mmaCell}[moredefined={DiracSimplify,FCVerbose,GSD,
GAD}]{Input}
  DiracSimplify[GAD[\(\mu\)].(GSD[p] + m).GAD[\(\mu\)], FCVerbose -> 3]
\end{mmaCell}
}
\noindent will not only return the final result but also generate a large amount of text output related to different stages of the evaluation inside the main function.

In total, there are 3 levels of verbosity, that differ in the amount of information
printed in the front end. Setting \texttt{FCVerbose} to 1 will merely show internal timings required to complete a particular operation. This can be helpful to localize potential bottlenecks in the code and possibly take countermeasures by changing the relevant options. For example,
if some function can quickly obtain the result but gets stuck while collecting the terms with respect to some symbols, one may check if setting the option \texttt{Collecting} (when available) to \texttt{False} can help to avoid this issue. With \texttt{FCVerbose->3}\footnote{\texttt{FCVerbose->2} as an intermediate level between 1 and 3 exists as well, but as of now it is not widely used. Very few functions also support \texttt{FCVerbose->4}, which was introduced during hunts for some tough bugs in the past.} the functions will also output intermediate results of the internal evaluations. This can significantly reduce the performance, as the front end would need to format the (possibly very large) intermediate expressions  accordingly. Therefore, this level of verbosity can be useful only when trying to fix a bug in the \feyncalc code or to understand some anomalous behavior of the package.

\subsection{Syntax checks}

When \feyncalc returns obviously wrong results, this may be caused not only by bugs in the code but also by syntax mistakes and inconsistencies in the input expressions. For example, the use of \texttt{Times} instead of \texttt{Dot} when writing down strings of Dirac or color matrices is a very common mistake among new users. Therefore, it may come as a surprise that in general \feyncalc happily accepts obviously incorrect inputs such as 
{\small
\begin{mmaCell}[moredefined={DiracSimplify,GA}]{Input}
  DiracSimplify[GA[\(\mu\)] GA[\(\nu\)] GA[\(\mu\)]]
\end{mmaCell}
}
\noindent or
{\small
\begin{mmaCell}[moredefined={Contract,FV,MT}]{Input}
  Contract[MT[\(\mu\), \(\nu\)] FV[p, \(\mu\)] FV[p, \(\mu\)]]
\end{mmaCell}
}
\noindent In the past, we received many requests to implement thorough syntax checks that would detect such errors, abort the evaluation and warn the user. Unfortunately, from our experience, such checks are extremely expensive performance-wise and can never achieve a 100\% success rate. Therefore, a syntax checker that by default analyzes the input of every high-level function is clearly not a viable solution.

Nonetheless, in order to address some of these issues in \feyncalc 9.3 we added a new function \texttt{FCCheckSyntax} that attempts to validate the given input by searching for the most common input errors, \eg
{\small
\begin{mmaCell}[moredefined={FCCheckSyntax,GA}]{Input}
  FCCheckSyntax[GA[\(\mu\)] GA[\(\nu\)] GA[\(\mu\)]]
\end{mmaCell}
\begin{mmaCell}{Message}
  FCCheckSyntax::failmsg: Error! FCCheckSyntax has found an inconsistency in your input expression and must abort the evaluation. The problem reads: Commutative products of DiracGamma in  \mmaSup{(\mmaSup{\mmaOver{\(\gamma\)}{_}}{\(\mu\)})}{2} \mmaSup{\mmaOver{\(\gamma\)}{_}}{\(\nu\)}
\end{mmaCell}
}
\noindent While \texttt{FCCheckSyntax} obviously cannot detect 
every possible inconsistency, we are convinced that it can be very helpful for beginners that are not yet fully familiar with the \feyncalc syntax or have little prior experience with \mma. When in doubt, one can
always apply \texttt{FCCheckSyntax} to the given expression and check if an error message is generated.

\section{Summary} \label{sec:summary}

In this article we presented a new version \feyncalc, a well-known \mma package for symbolic calculations in QFT and particle physics. Some new features of \feyncalc 9.3 that should be interesting for a broad audience of 
practitioners include a practical solution to shadowing issues, native ability to determine the UV-divergent pieces of arbitrary Passarino--Veltman functions, support for calculations with Majorana spinors, introduction
of Dirac matrices and spinors with explicit Dirac indices and automatic determination of relative signs for 4-fermion vertices in amplitudes generated with \feynarts. Furthermore, in order to promote good practices of employing \feyncalc in loop calculations, we tried to address some questions that may arise in the daily use of the package and to warn the users against potential pitfalls.

In order to put the amount of new features and the time passed since the previous stable release 
into perspective, we would like to state that most changes in the code of \feyncalc between versions 9.2 and 9.3 were related to the still experimental support for Cartesian tensors, Pauli matrices and nonstandard loop integrals that are crucial for making \feyncalc applicable to nonrelativistic calculations.  As the corresponding symbols and routines are not yet finalized, we restricted ourselves to briefly mentioning their existence and referring interested readers to the existing documentation.

To summarize, we are convinced that \feyncalc can be a very handy tool in the software toolkit of every HEP phenomenologist, be it for professional or educational use. The enormous flexibility in choosing the way how to organize and carry out a particular calculation, many worked out examples and a large, friendly community make \feyncalc stand out among similar software packages and contribute to its popularity in the particle physics community. 

The developers of the package are strongly committed to continue working towards a faster and more feature-rich \feyncalc, with a special emphasis on EFTs and multiloop calculations. The former should be addressed in the \textsc{FeynOnium} project, while for the latter we plan to add a native interface to \texttt{QGRAF}~\cite{Nogueira:1991ex} and a set of routines for topology identification and better interoperability with \textsc{FORM} along the lines of the approaches used in~\cite{Feng:2012tk} and~\cite{Cyrol:2016zqb}.

We are also investigating possibilities to provide a completely web-based version of \feyncalc that would run on a public and private \textsc{Wolfram Enterprise Cloud}. This could be useful not only as a showcase of \feyncalc's abilities (\eg by reproducing known results from the literature) but also for educational purposes.

\section*{Acknowledgments}

We thank Georg Sulyok for providing us with an improved version of his algorithm for determining UV-divergent parts of one-loop tensor integrals in dimensional regularization~\cite{Sulyok:2006xp} and giving his permission to integrate it into \feyncalc.

We are indebted to all \feyncalc users who actively followed the development of the version 9.3 in the past years by reporting the encountered bugs (including nonobvious bugs in \textsc{TARCER} \eg as discovered in~\cite{Blumlein:2011mi}), regressions and inconsistencies or submitting patches to the repository. A special gratitude goes to Aliaksei Kachanovich, who carefully tested the development version of \feyncalc in his daily research~\cite{Kachanovich:2020xyg,Kachanovich:2020yhi} for almost two years and provided regular, detailed feedback on encountered issues, performance bottlenecks, unclear documentation and various user-experience annoyances.

One of the authors (V.\,S.) would like to acknowledge Nora Brambilla, Antonio Vairo, Thomas Hahn, Martin Beneke, Sergey Larin, Jos Vermaseren, William J. Torres Bobadilla, Hirten Patel, Gudrun Heinrich, Hua Xing Zhu, Tong-Zhi Yang, Sven-Olaf Moch, Oleksandr Gituliar, Matthias Steinhauser, Ulrich Nierste, Alexander Smirnov, Vladimir Smir\-nov, Konstantin Chetyrkin, Florian Herren and Marvin Gerlach for useful discussions on different aspects of automatic perturbative calculations. He would also like to thank Instituto de F\'isica Corpuscular (IFIC) in Valencia,
Technical University of Munich (TUM) and the organizers of the 13th International Workshop on Heavy Quarkonium (QWG 2019) in Turin, were parts of this work were done, for hospitality.

The research of V.\,S. was supported  by  the  Deutsche Forschungsgemeinschaft  (DFG,  German Research Foundation) under grant 396021762 - TRR 257 ``Particle Physics Phenomenology after the Higgs Discovery'', National Science Foundation of China (11135006, 11275168, 11422544, 11375151, 11535002) and the Zhejiang University Fundamental Research Funds for the Central Universities (2017QNA3007). We equally acknowledge the support from the DFG and the NSFC through funds provided to the Sino-German CRC 110 ``Symmetries and the Emergence of Structure in QCD'' (NSFC Grant No. 11261130311) and from the DFG cluster of excellence ``Origin and structure of the universe'' (\url{www.universe-cluster.de}). Additional support was provided  by the Munich Institute for Astro- and Particle Physics (MIAPP) of the DFG cluster of excellence ``Origin and Structure of the Universe''.

\appendix

\section{Experimental support for new propagators and tensor types}

It is worth mentioning that \feyncalc 9.3 also contains many new features and capabilities 
that are still considered not sufficiently stable or even experimental. They are mainly related to the native support for Cartesian tensors, Pauli algebra and loop integrals with nonstandard propagators. The main goal of these efforts is to make \feyncalc more useful for EFT calculations, with a special focus on EFTs of strong interactions and nonrelativistic EFTs (NREFTs). Although all the corresponding symbols and functions are properly documented and available to the user of the version 9.3, their syntax is still subject to change in the future. It will be finalized with the official release of the \textsc{FeynOnium}\footnote{The original idea for this project arose during the work on~\cite{Brambilla:2017kgw}. Development versions of \feyncalc and \textsc{FeynOnium} were then employed in~\cite{Brambilla:2019fmu}.} add-on, that will be described elsewhere~\cite{Brambilla:2020fla}.

To provide a sneak preview of what is already possible with \feyncalc 9.3 in this respect, let us give one example of a loop integral manipulation that was completely out of reach in the older versions of the package. Using the new representation of denominators \texttt{Standard\-Feyn\-Amp\-Denominator} (abbreviated as \texttt{SFAD}) we can directly write down an integral with eikonal propagators and simplify it \eg by applying partial fractioning 
{\small
\begin{mmaCell}[moredefined={SFAD,ApartFF,FCMultiLoopTID,SPD}]{Input}
  SFAD[\{\{0, 2 p.q\}\}, p + q, q]
  ApartFF[\%, \{q\}]
\end{mmaCell}
{ \small
\begin{mmaCell}{Output}
  \mmaFrac{1}{(2\,(p\(\cdot\)q)+i\(\eta\)))(\mmaSup{(p+q)+i\(\eta\))}{2}.(\mmaSup{q}{2}+i\(\eta\))}
\end{mmaCell}
}
{ \small
\begin{mmaCell}{Output}
  -\mmaFrac{1}{\mmaSup{p}{2}(\mmaSup{q}{2}+i\(\eta\)).(\mmaSup{(p+q)+i\(\eta\))}{2}} + \mmaFrac{1}{\mmaSup{p}{2}(2\,(p\(\cdot\)q)+i\(\eta\))).(\mmaSup{q}{2}+i\(\eta\))} - \mmaFrac{1}{\mmaSup{p}{2}(2\,(p\(\cdot\)q)+i\(\eta\))).(\mmaSup{(p+q)}{2}+i\(\eta\))}
\end{mmaCell}
}
}
There also exists a Cartesian version of \texttt{SFAD} called \texttt{CFAD}. Tensor reductions of Cartesian integrals with usual propagators are possible using \texttt{TID}, while \texttt{FCMultiLoopTID} can process tensor eikonal integrals which often cannot be reduced to scalar integrals with unit numerators. Nonetheless, the treatment of such integrals within \feyncalc is still very much work in progress and we hope to improve on that in the subsequent versions of the package.

\bibliographystyle{h-physrev}
\bibliography{biblio.bib}

\end{document}